\documentclass[twocolumn]{aastex631}
\DeclareUnicodeCharacter{2212}{\ensuremath{-}}

\usepackage{float}
\graphicspath{{./}{figures/}}
\shorttitle{Neutrinos in IceCube and KM3NeT}

\begin{document}

\title{\textbf{Neutrinos in IceCube and KM3NeT}}

\email{shiqi.yu@utah.edu}
\author{Shiqi Yu}
\affiliation{Department of Physics and Astronomy, University of Utah, Salt Lake City, UT 84112, USA}

\begin{abstract}
%250 wd limit
Neutrino observatories such as IceCube, Cubic Kilometre Neutrino Telescope (KM3NeT), and Super-Kamiokande cover a broad energy range that enables the study of both atmospheric neutrinos and astrophysical neutrinos. IceCube and KM3NeT focus on a similar energy range, from a few GeV to PeV, and have conducted competitive work on the atmospheric neutrino flux, three-flavor oscillation parameter measurements, searches beyond the Standard Model, and investigations of cosmic-ray accelerators using high-energy astrophysical neutrinos. Recent IceCube findings of evidence of neutrino signals from NGC~1068 have triggered a series of follow-up studies. These studies provide evidence that a subset of Seyfert galaxies may produce high-energy neutrinos. The emerging candidates are NGC~4151, NGC~3079, CGCG~420-015, and Circinus Galaxy. 
Furthermore, a stacking analysis of 13 selected sources in the Southern Hemisphere reported a cumulative neutrino signal at 3.0\,$\sigma$, offering independent evidence that some X-ray-bright Seyfert galaxies could be potential high-energy neutrino sources.
KM3NeT, still under construction, continues to accumulate data that will support future studies of astrophysical neutrino sources. However, with its currently deployed detection units, it has detected an ultra-high-energy event of several tens of PeV originating from approximately 1 degree above the horizon.
This contribution highlights and summarizes recent findings from IceCube and KM3NeT in both neutrino physics and astrophysics.
% 1 sentence about KM3NeT hightE neutrino event
\end{abstract}
%\bigskip
%\vspace{5cm}% Additional space between abstract & rest of document

%\keywords{%editorials, notices --- miscellaneous --- catalogs --- surveys}

\section{Introduction}
High-energy neutrinos can be produced in the atmosphere or through high-energy processes in the universe. Neutrinos produced in the atmosphere can be used to study neutrino oscillations, while high-energy astrophysical neutrinos serve as unique messengers, providing a window to the universe's most energetic phenomena -- cosmic-ray accelerators.
%Neutrino telescopes such like IceCube and KM3NeT can study both. 

The 2015 Nobel Prize in Physics was awarded for the discovery of neutrino oscillations, a groundbreaking result that confirmed neutrinos have mass~\citep{PhysRevLett.81.1562,PhysRevLett.87.071301}. %This finding provided crucial evidence for the phenomenon of neutrino flavor oscillation, where neutrinos transition between different flavors--electron, muon, and tau--as they propagate. 
It challenged the Standard Model of particle physics and expanded our understanding of fundamental particle behavior, opening the door to new explorations in physics and cosmology.

Neutrinos can also serve as valuable messengers for studying distant objects in the universe due to their proprieties--nearly massless and interacting only weakly with matter. The first direct detection of high-energy astrophysical neutrinos by IceCube in 2013 opened a new window for us to study the universe, confirming the existence of a high-energy neutrino flux from astrophysical sources~\citep{IceCube2013}. Additionally, one of IceCube’s recent milestones~\citep{2022}, the evidence of detecting neutrino emission from the nearby active galactic nucleus, NGC~1068 (a type II Seyfert galaxy), has provided key insights into the nature of cosmic-ray accelerators.
%  \begin{figure}[tbh!]
%\includegraphics[trim={15cm 40 0 40},clip,width=\linewidth]{plots/detector.jpeg}
%\caption{Sketch of IceCube detector with IceCube-DeepCore marked in light-green shade.}\label{fig:ic}
%  \end{figure}
IceCube, located at the South Pole, deep under the Antarctica ice, consists of 86 strings and is made of 5,160 digital optical modules (DOMs). 
IceCube-DeepCore (or DeepCore), a sub-detector, is located at the bottom center of the main array.
%, as shown in the green-shaded region in Fig.~\ref{fig:ic}. 
The DOMs on the DeepCore strings are more closely spaced and have higher quantum efficiency than those on the IceCube strings, which lowers the detection threshold to a few GeV and enables the study of atmospheric neutrino oscillations.
KM3NeT is located at the bottom of the Mediterranean Sea and operates across two sites: ORCA, the low-energy subdetector situated off the coast south of Toulon, France, and ARCA, the high-energy subdetector, located off the coast of Sicily, Italy.
%TODO add a building block here
ORCA is planned to consist of a single building block, while ARCA is designed to comprise two building blocks. Each block houses 115 detection units (DUs), each equipped with 18 digital optical modules (DOMs). In ARCA, the spacing between DOMs and DUs is greater than in ORCA, optimizing it for the detection of high-energy neutrinos.

IceCube (the main array) and ARCA are designed to study high-energy astrophysical neutrinos, while DeepCore and ORCA focus on atmospheric neutrino oscillations. Both ARCA and ORCA are currently under construction, with 28 and 23 detection units (DUs) deployed, respectively, as of the latest update~\cite{KM3NeT:nu2024}.
KM3NeT’s location provides strong sensitivity to sources in the Southern Hemisphere, complementing IceCube's optimal sensitivity to sources in the Northern Hemisphere.

\section{Atmospheric Neutrinos}

Precisely measuring oscillation parameters is crucial for understanding neutrinos and advancing research into physics beyond the Standard Model. Atmospheric neutrinos are especially valuable for studying neutrino oscillations. Both KM3NeT and IceCube have made significant contributions to this field. Key results from both experiments in the study of atmospheric neutrinos and neutrino oscillations are discussed in this section.

%Atmospheric neutrinos are produced when cosmic rays interact with atmospheric nuclei, with pion and kaon decays dominating the production process. This leads to a higher flux of neutrinos compared to anti-neutrinos.

\subsection{ATMOSPHERIC NEUTRINO FLUX MEASUREMENT}

Atmospheric neutrinos are produced when cosmic rays interact with atmospheric nuclei, with the decay of pions and kaons dominating the production process. Neutrino observatories, such as AMANDA-II~\citep{Abbasi_2009}, IceCube~\citep{Aartsen_2017}, ANTARES~\citep{Adrian-Martinez:2013}, and Super-Kamiokande~\citep{Richard_2016},  can directly measure the atmospheric neutrino flux.

\begin{figure}[tbh!]
    \centering
    \includegraphics[width=\linewidth]{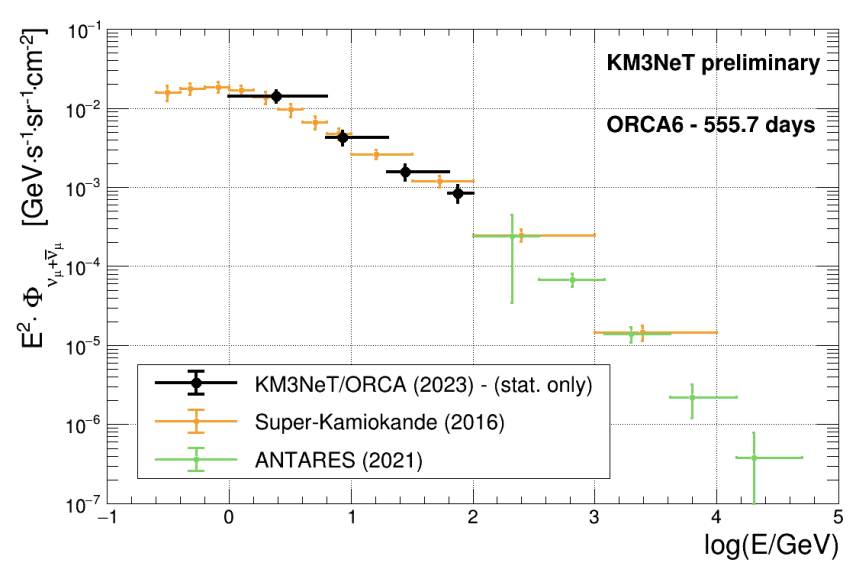}
    \caption{Atmospheric muon neutrino flux measurement using data taken with ORCA6 data (black) compared with the measurements from Super-Kamiokande (orange,~\cite{Richard_2016}) and ANTARES (green,~\cite{Adrian-Martinez:2013}).}
    \label{fig:orca_atmflux}
\end{figure}

A more recent study was conducted by KM3NeT using the ORCA6 configuration (first 6 DUs, with 555.7 days of data), where~\cite{Stavropoulos:2023FC} reported a preliminary measurement of the atmospheric muon neutrino flux in the range of 1–100 GeV, as shown in Fig.~\ref{fig:orca_atmflux}. 
This result shows excellent agreement with the Super-Kamiokande result in~\cite{Richard_2016} below 100 GeV, while also extending the lower-energy range of the findings from KM3NeT’s predecessor, ANTARES~\citep{Adrian-Martinez:2013}.

\subsection{MEASUREMENTS OF NEUTRINO OSCILLATIONS}
\begin{figure}[tbh!]
\includegraphics[width=\linewidth]{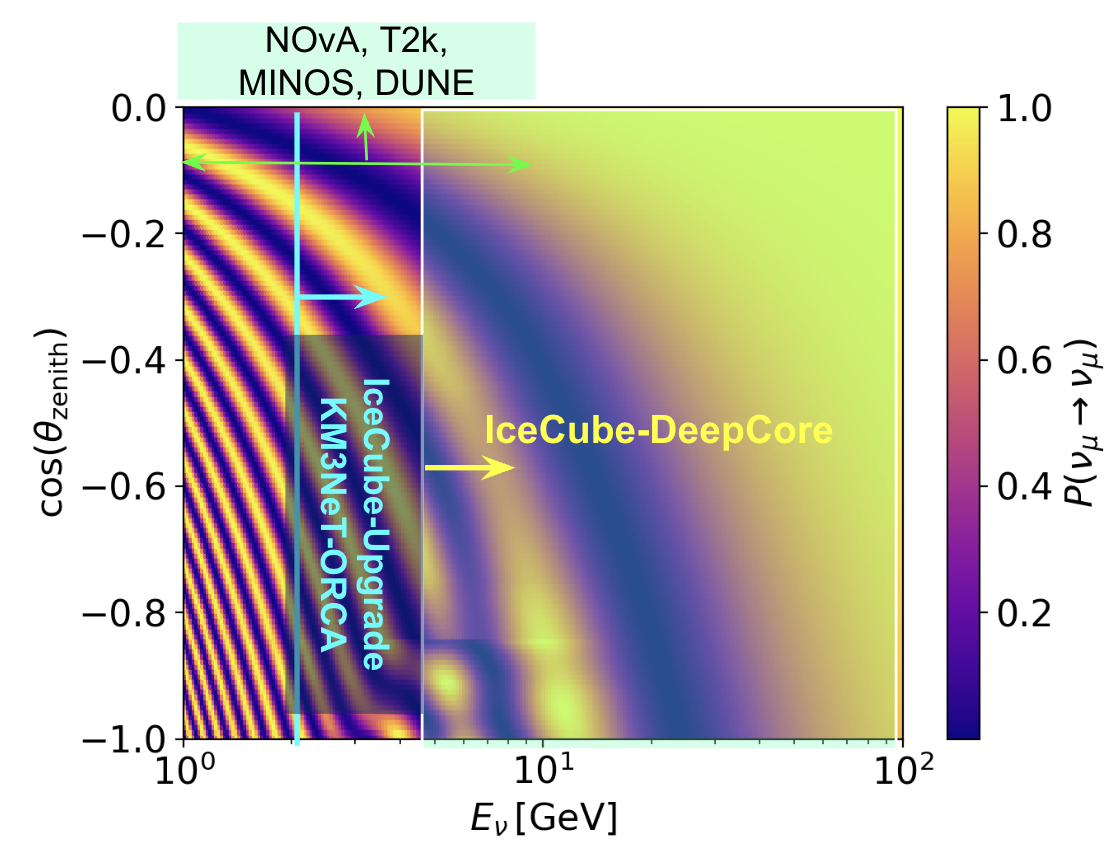}
\caption{The muon neutrino survival probability (color), assuming normal ordering of neutrino masses, is plotted as a function of true $\cos(\theta_{\rm{zenith}})$ versus true energy. Sketched arrows and shaded regions indicate the visibility of DeepCore (yellow), IceCube-Upgrade and KM3NeT-ORCA (teal), and long-baseline accelerator-based experiments (green).}
\label{fig:osc}
\end{figure}

Neutrino oscillations depend on both the neutrino energy ($E$) and the distance traveled ($L$). For atmospheric neutrinos, $L$ can be inferred from the neutrino's arrival direction, $\cos(\theta_{\rm{zenith}})$. Unlike long-baseline accelerator-based neutrino experiments, which typically use fixed baselines and rely on fitting a 1D oscillation probability curve, atmospheric neutrinos cover a wide range of baselines and energies, as illustrated in the 2D oscillogram in Fig.~\ref{fig:osc}.
The positions of the oscillation valleys in the 2D oscillogram provide strong constraints on the measurement of $\Delta m^2_{32}$.

\begin{figure}[tbh!]
\centering\includegraphics[width=\linewidth]{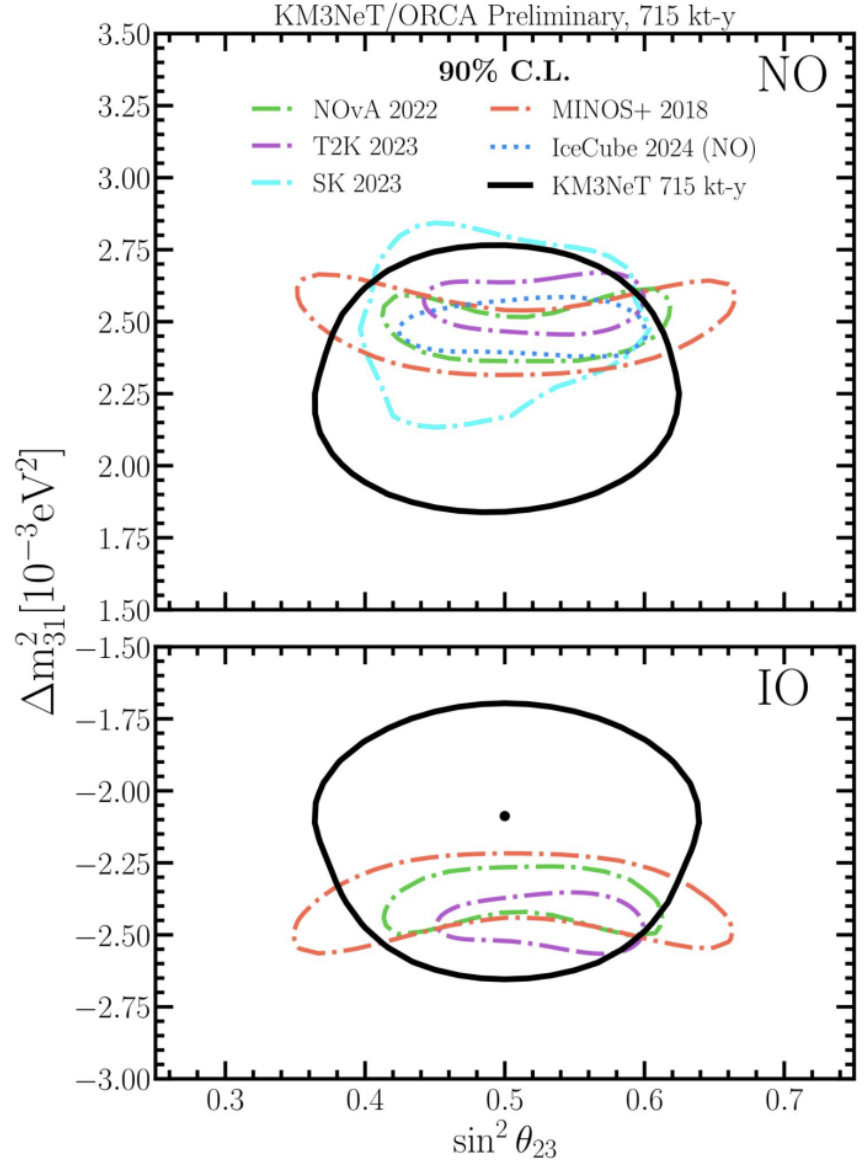}
\caption{Allowed regions at 90\% CL for the oscillation parameters \( \sin^2 \theta_{23} \) and \( \Delta m_{31}^2 \) are shown for KM3NeT (black,~\cite{KM3NeT:nu2024}) in both NO and IO, and for IceCube (dotted,~\cite{IceCubeDeepCore2024}) in NO only, compared to results from other experiments. Plot is from~\cite{KM3NeT:nu2024}.}\label{fig:numu_osc}
\end{figure}

KM3NeT reported measurements of the muon neutrino disappearance ($\nu_\mu \rightarrow \nu_\mu$) probability under both normal (NO) and inverted (IO) neutrino mass ordering assumptions, as shown in Fig.~\ref{fig:numu_osc} from~\cite{KM3NeT:nu2024}. These results were obtained using the ORCA11 configuration with 715 days of data, which is equivalent to 37 days of data collected with the full ORCA detector.
The ORCA11 sample consists of approximately 10,000 high-purity track-like events, reconstructed using a combination of boosted decision tree and likelihood techniques, as detailed in~\cite{KM3NeT_ORCA2024}. This result is consistent with measurements from other experiments, with a slight preference for inverted ordering (IO) at $2\ln(L_{\rm IO}/L_{\rm NO}) = 0.61$, as reported in~\cite{KM3NeT:nu2024}.

The DeepCore sample, enhanced by the use of convolutional neural networks for reconstruction and event selection, consists of over 150,000 selected events accumulated over 9.3 years of data. With this large dataset, IceCube provided a competitive measurement of the $\nu_{\mu}$ disappearance parameters and their 1\,$\sigma$ uncertainties in NO~\cite{IceCubeDeepCore2024}: $\Delta m^2_{32} = 2.40^{+0.05}_{-0.04} \times 10^{-3}\ \textrm{eV}^2$ and $\sin^2\theta_{23} = 0.54^{+0.04}_{-0.03}$.

\begin{figure}[tbh!]
    \centering    \includegraphics[width=\linewidth]{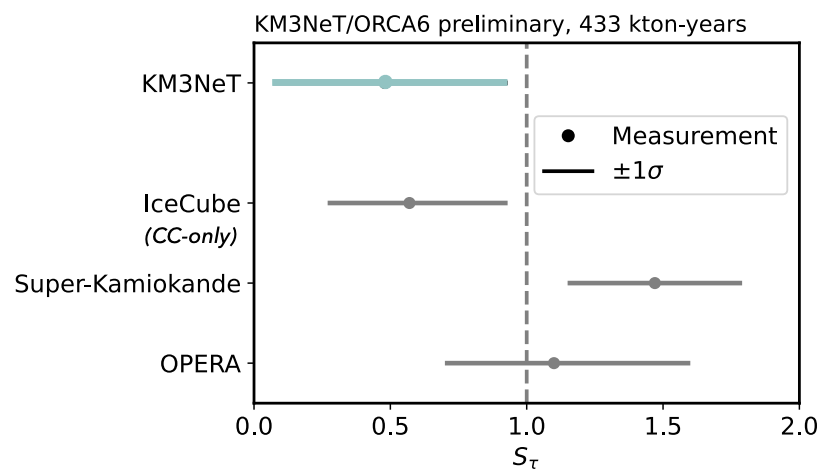}
    \caption{Atmospheric $\nu_\tau$ appearance measurements using data from ORCA6 (green), compared with results from other experiments~\citep{Aartsen2019,superk_tau_appearance_2018,Agafonova2018}.}
    \label{fig:tau}
\end{figure}

\begin{figure}[tbh!]
    \centering
\includegraphics[width=\linewidth]{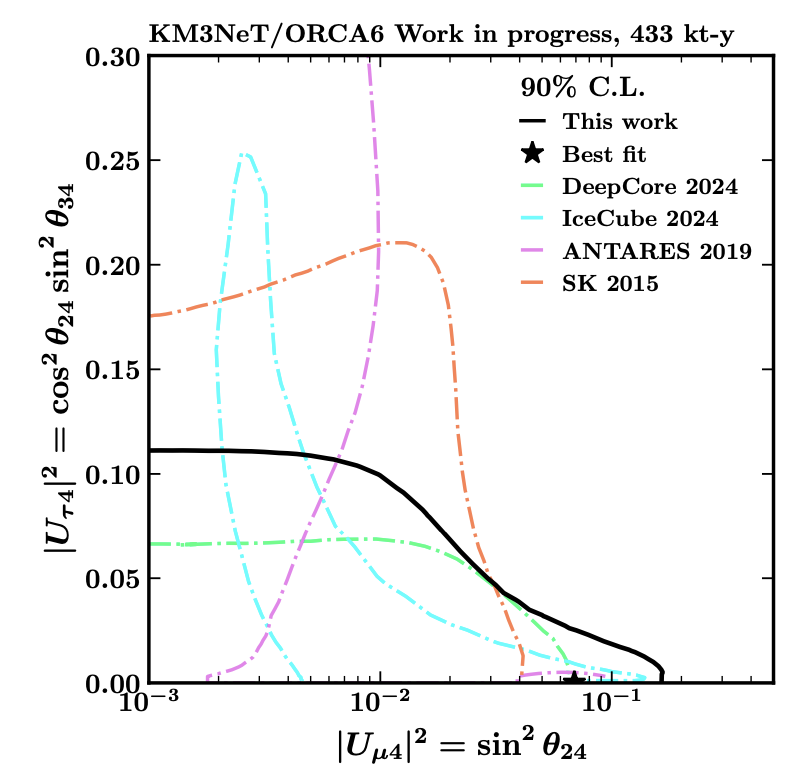}
    \caption{Contour of the 90\% CL limit of measurements from ORCA (black), DeepCore (green,~\cite{sterile_verification}), IceCube TeV study (teal,~\cite{tevsterileIceCube2024}), ANTARES (pink,~\cite{ANTARES2019}), and Super-Kamiokande (orange, ~\cite{PhysRevD.91.052019}). }
    \label{fig:sterile}
\end{figure}

In addition to the atmospheric $\nu_\mu$ disappearance measurement, KM3NeT, under the ORCA6 configuration, also studied the $\nu_\mu \rightarrow \nu_\tau$ channel to measure the normalization of $\nu_\tau$ charged-current interactions~\citep{KM3NeT:tau}. The reported precision is remarkably comparable to that of other experiments~\citep{Aartsen2019,superk_tau_appearance_2018,Agafonova2018}, as shown in Fig.~\ref{fig:tau}.
Beyond the three-flavor framework, neutrino observatories have also searched for sterile neutrinos in both the GeV and TeV energy ranges. As shown in Fig.~\ref{fig:sterile}, 
IceCube and KM3NeT reported compatible constraints using updated sub-100 GeV sample in~\cite{sterile_verification} and~\cite{KM3NeT:sterile}, respectively. Additionally, utilizing a TeV muon-neutrino-dominated sample, IceCube conducted a study~\citep{tevsterileIceCube2024}, with results in strong agreement with other measurements, as shown in Fig.~\ref{fig:sterile}. Although TeV-range studies exceed ORCA's energy capabilities, ARCA presents an exciting opportunity for future exploration in this direction.

While IceCube has made significant progress in narrowing the parameter ranges, KM3NeT is still in the process of construction and accumulating statistics. Both experiments face challenges related to their Monte Carlo (MC) models, which could benefit from mutual improvements and collaboration. The extensive dataset accumulated by IceCube over the past decade has substantially reduced the impact of statistical uncertainty, in contrast to the uncertainties stemming from the MC models.
Meanwhile, KM3NeT continues to face challenges from limited statistics and energy scale uncertainties, which remain the largest systematic uncertainty, primarily driven by uncertainties in water properties~\citep{KM3NeT_ORCA2024}.

\section{Astrophysical Neutrinos}
The primary scientific goal of large-scale neutrino observatories like IceCube and KM3NeT is to study high-energy astrophysical neutrinos and their origins. Recent findings on high-energy astrophysical neutrinos from NGC~1068 have drawn attention to X-ray-bright Seyfert galaxies. During cosmic-ray acceleration and interaction, multi-messenger signals—such as multi-wavelength photons and neutrinos—are produced. Studying these signals can provide valuable insights into the nature of large-scale cosmic-ray accelerators in the distant universe. High-energy neutrinos are believed to originate from hadronic processes, where high-energy cosmic rays interact with protons or gamma rays.  
%Independent analyses have been conducted to identify similar sources, employing different theoretical assumptions.
%These high-energy neutrinos are believed to originate from hadronic processes, where high-energy cosmic rays interact with protons or gamma rays. 

Independent searches have identified X-ray-bright Seyfert galaxies as potential sources of cosmic neutrinos. 
In~\cite{IceCube:2024dou}, a search shows an excess of the collective neutrino emission from a catalog of X-ray-bright Seyfert galaxies in the Northern Hemisphere, with a significance of 2.7\,$\sigma$, driven primarily by contributions from NGC~4151 and CGCG~420-015. 
Additionally, IceCube reports a 2.9\,$\sigma$ (post-trial) neutrino excess from NGC~4151 in~\cite{hardxray}. In ~\cite{Neronov:2023aks}, neutrino excesses from both NGC~4151 and NGC~3079 are reported, with a combined p-value for the coincident detection of neutrino excesses from these two sources of less than $2.6\times10^{-7}$, corresponding to a 5\,$\sigma$ significance.
A recent analysis of Seyfert galaxies in the Southern Hemisphere, presented in~\cite{circinus}, provides independent evidence of neutrino emission from a collective X-ray-bright Seyfert galaxies, potentially originating from the Circinus Galaxy, ESO~138-1, and NGC~7582.

These findings underscore the growing capability of neutrino observatories to identify and study cosmic neutrino sources.
%In addition to the studies focused on sources in the Northern sky (with declination angles greater than -5 degrees), t

The study in~\cite{circinus}, which focused on sources in the Southern Hemisphere, employed advanced analysis techniques to mitigate contamination from high-energy neutrinos produced within the Milky Way~\cite{2023gp}. Galactic neutrinos have posed a significant challenge when studying point-like sources near the Galactic Plane, such as the Circinus Galaxy and Centaurus A. The improved techniques have successfully addressed this challenge by accurately accounting for galactic neutrinos in the background modeling of the sources under study. This advancement not only benefits the study in~\cite{circinus} but also paves the way for future work targeting sources near the Galactic Plane.

In addition to studying individual galaxies,~\cite{circinus} performed a search for the cumulative neutrino signals from the selected sources using a stacking method~\citep{Achterberg_2006}. In the stacking analysis, each source was weighted according to the expected number of signal events detectable by IceCube, which was estimated using the neutrino spectrum derived from the disc-corona model~\citep{Kheirandish:2021wkm}.

\begin{figure}[t!]
\centering
\includegraphics[width=\linewidth]{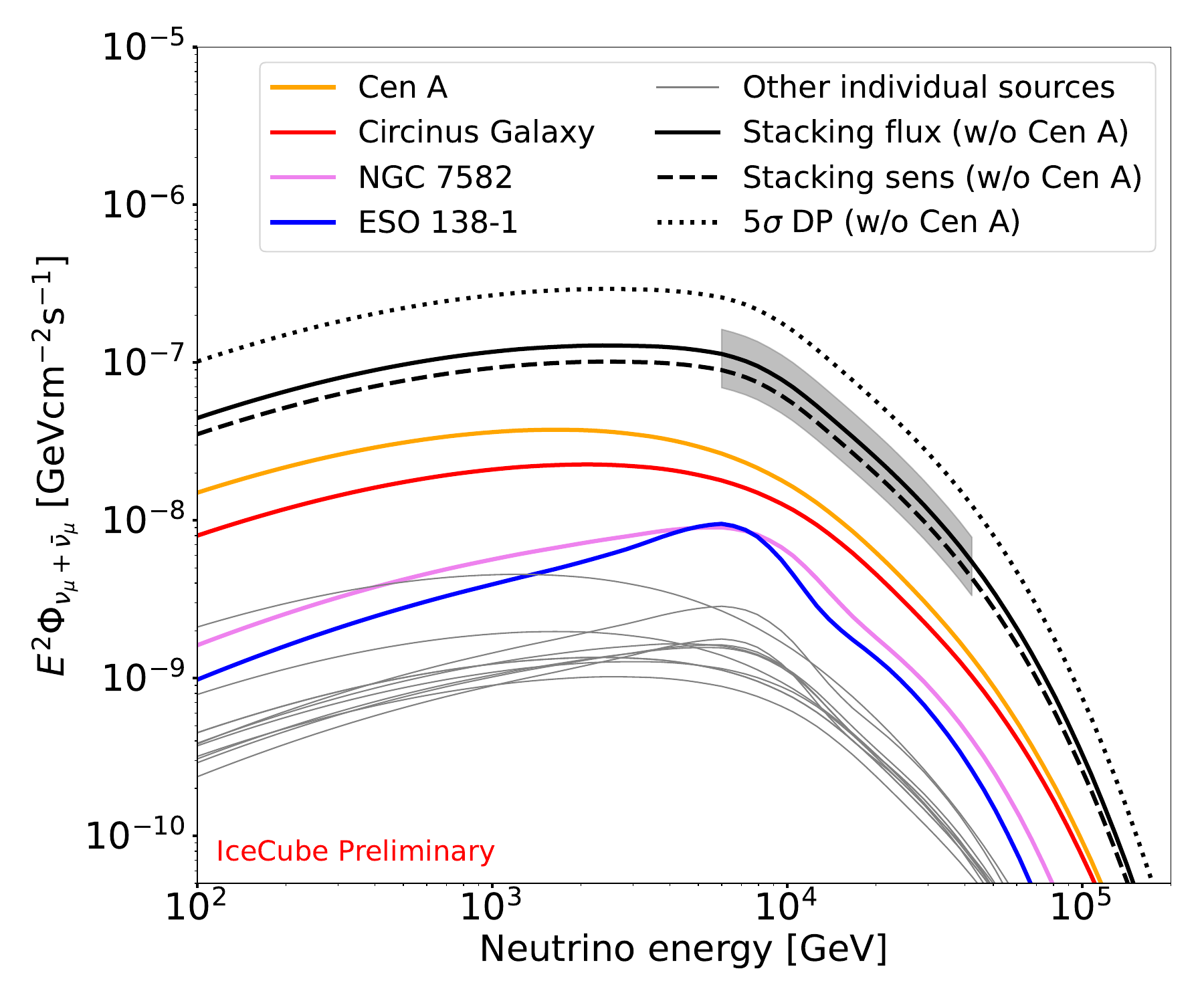}
\caption{Expected energy spectrum of each source (gray) from the disc-corona model, with the top 4 sources highlighted in color. The total stacking flux (excluding Centaurus A) at the best-fit is shown as a solid black curve, with 1\,$\sigma$ errors indicated by the gray shaded band, and the 5\,$\sigma$ discovery potential is represented by the dotted curve.}
    \label{fig:stack}
\end{figure}
The stacking search identified a combined excess of 6.7 events from 13 selected sources, which is inconsistent with the background hypothesis at a significance of $3.0\sigma$, as shown in Fig.~\ref{fig:stack}. This provides independent evidence, complementing previous findings, that some X-ray-bright Seyfert galaxies may be sources of high-energy neutrinos. A summary of the results is provided in Table~\ref{tab:results}.

\begin{figure}[tbh!]
\centering
\includegraphics[width=\linewidth]{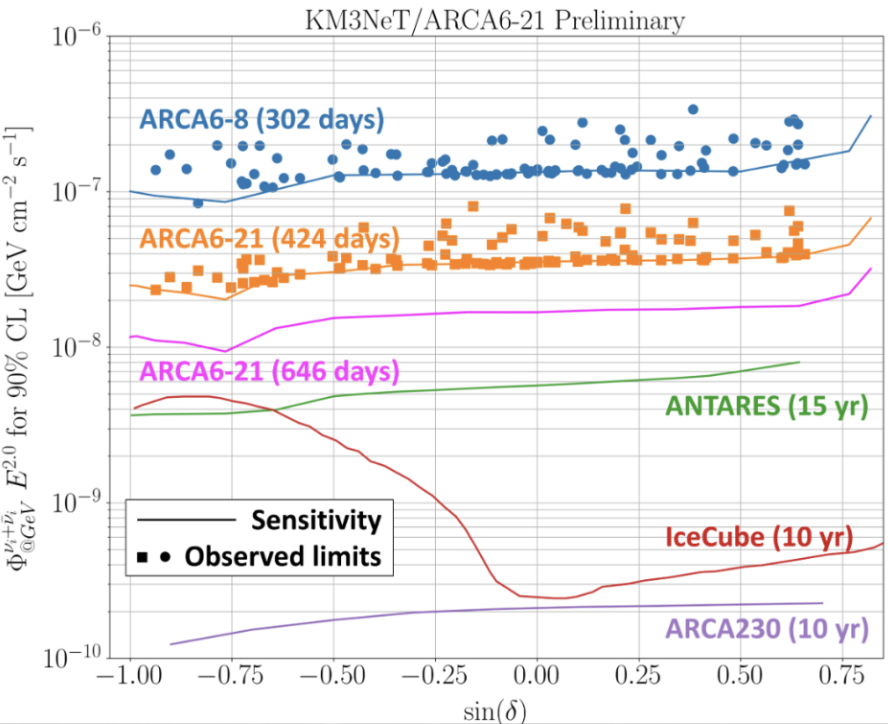}
\caption{Comparison of the observed limits on the flux for the ARCA6-21 point source analysis assuming E$^{-2}$ spectrum of full sky, with ANTARES (15-year,~\cite{alves2023antares}) and IceCube (10-year,~\cite{PhysRevLett.124.051103}) sensitivities.}
\label{fig:arca_sens}
\end{figure}

KM3NeT’s ARCA is still under construction, with its sensitivity improving over time and approaching that of ANTARES, as shown in Fig.~\ref{fig:arca_sens}. 
Once completed, ARCA is expected to offer better sensitivity in the Southern Hemisphere than IceCube, as detailed in~\cite{km3net2024arca, km3net2024astronomy}.
Despite being under construction, KM3NeT reported the detection of an exceptionally high-energy event (several tens of PeV) in~\cite{KM3NeT:nu2024}. This ultra-high-energy neutrino, which arrived from approximately 1 degree above the horizon, is an outlier--occurring in only 1 in a million detected events by ARCA so far--and illuminated 3,672 photomultiplier tubes, corresponding to 35\% of the total detector array.

\begin{deluxetable*}{lDDDDD}\label{tab:results}
\tablecaption{Summary of the most significant sources from IceCube searches and the stacking result using the disc-corona model~\citep{Kheirandish:2021wkm}. The table lists the best-fit number of signals ($\hat{n}_{\rm s}$), spectral index ($\hat{\gamma}$), pre-trial p-value ($p{\rm local}$), best-fit flux ($\phi^{1 \rm{TeV}}_{\nu_{\mu}+\bar{\nu}_{\mu}}$), and the 90\% confidence level upper limit for all analysis results. Upper limits per flavor are estimated at a normalization energy of 1\,TeV with a flux spectral index of $E^{-3}$, in units of $\times10^{-13}{\rm TeV}^{-1}{\rm cm}^{-2}{\rm s}^{-1}$ for the power-law analysis. For model-based analyses, the expected number of signal events ($n_{\rm exp}$) is listed.}
\tablewidth{\textwidth}
\tablehead{
\colhead{} & \multicolumn2c{$\hat{n}_{\rm s}$} & \multicolumn2c{$\hat{\gamma}$} & \multicolumn2c{$p_{\rm local}$} & \multicolumn2c{$\phi^{1 TeV}_{\nu_{\mu}+\bar{\nu}_{\mu}}$} &\multicolumn2c{90\% U.L.} %& 
}
\decimals
\startdata
NGC~$1068^1$ &81.7 & 3.1 & $1.27\times10^{−6}$\,(4.7\,$\sigma$) & $4.02\times10^{-11}$ & $-$ \\
NGC~$4151^1$ &49.8 & 2.83 & $3.99\times10^{−5}$\,(3.9\,$\sigma$) & $1.51\times10^{-11}$ & $-$  \\
NGC~3079$^1$ &29.53 &4.0 &0.003\,(2.7\,$\sigma$) & $-$ & 203.50 \\
CGCG~420-015$^2$ &35 &2.8 &0.003\,(2.7\,$\sigma$) & $-$ & 25.9 \\
Circinus ${\rm Galaxy}^3$ &3.1 &2.5 &0.001\,(3.1\,$\sigma$) & $-$ & 63.80 \\
\hline
Stacked sources & \phantom{$-$} &  \phantom{$-$} & \phantom{$-$} & \hspace{0.1cm} $n_{\rm exp}$  & \hspace{0.1cm} $n_{\rm event}$ \\ \hline
13 sources (Southern sky)$^3$  & 6.7 & $-$ & 0.0013\,(3.0\,$\sigma$) &  4.7 &  14.3 
\enddata
\tablenotetext{1}{~\cite{hardxray},  $^2$~\cite{IceCube:2024dou}, $^3$~\cite{circinus}}
\end{deluxetable*}

\section{SUMMARY}
Neutrino oscillation studies using atmospheric neutrinos benefit from the broad energy and baseline ranges provided by neutrino observatories, offering strong constraints on the mass-squared difference ($\Delta m^2_{32}$). Both IceCube-DeepCore and KM3NeT-ORCA have reported new results from their atmospheric neutrino studies. IceCube has been collecting data with its full detector since 2011, providing a stable and extensive dataset, while KM3NeT-ORCA has been accumulating high-quality data during its construction phase. The next generation of atmospheric neutrino detectors, including IceCube-Upgrade, a fully configured ORCA, JUNO~\citep{juno2022}, and DUNE~\citep{abi2020deepundergroundneutrinoexperiment}, are on the horizon. These detectors have the potential to enable joint analyses or global fits that could lead to a $3\sigma$ determination of the neutrino mass ordering in the near future~\citep{junonmo}.

On the astrophysics side, studies have been conducted to follow up on the evidence of high-energy neutrino emission from NGC~1068. A recent analysis of selected X-ray-bright Seyfert galaxies in the Southern Hemisphere revealed a $3\sigma$ excess of accumulated neutrino emission from the stacked sources. This provides independent evidence that some X-ray-bright Seyfert galaxies could be potential candidates for high-energy neutrino production and, therefore, cosmic-ray acceleration. 
Once fully deployed, KM3NeT could significantly enhance searches in the Southern Hemisphere with its better sensitivity.

Together, IceCube and KM3NeT will provide full-sky coverage for observing astrophysical neutrinos and probing their origins. Furthermore, the integration of multi-messenger data and next-generation neutrino observatories—such as P-ONE~\citep{Twagirayezu:2023cpv}, TRIDENT~\citep{ye2024multicubickilometreneutrinotelescopewestern}, and HUNT~\citep{hunt}—holds great promise for identifying additional astrophysical neutrino sources and advancing our understanding of cosmic-ray accelerators.

\section*{Acknowledgements}
I would like to thank Alfonso Garcia, Andrii Terliuk, Joao Coelho, Victor Carretero, Dimitris Stavropoulos, Anuj Kumar, Zelimir Djurcic, Maury Goodman, Carsten Rott, Hermann Kolanoski, Hans Niederhausen, Chris Weaver, Zuan Ge, and many others for their valuable input, insightful discussions, and assistance in shaping this presentation, as well as for proofreading the manuscript. I also gratefully acknowledge the support and resources provided by the Center for High-Performance Computing at the University of Utah.

\bibliography{main}

\end{document}